\documentclass[aps,singlecolumn, showpacs,showkeys,secnumarabic,amssymb,amsmath,superscriptaddress]{revtex4-1}
\usepackage{amssymb,amsmath}
\usepackage{graphics}
\usepackage{graphicx}
\usepackage{bm}
\usepackage{color}
\usepackage{float}
\begin{document}
%
%
%
%

\title{Prediction of mass of $\eta_{c}(2S)$ using variational method}

\author{Chandrashekar Radhakrishnan}
\email{chandrashekar10@gmail.com}
\affiliation{Laboratoire Syst{\`e}mes Complexes et Information Quantique, ESIEA Group, 9 Rue V{\`e}sale, Paris 75005, France}
\affiliation{New York University Shanghai, 1555 Century Avenue, Pudong, Shanghai 200122, China.}
\affiliation{NYU-ECNU Institute of Physics at NYU Shanghai, 3663 Zhongshan Road North, Shanghai, 200062, China.}

\begin{abstract}
The suitability of using non-relativistic quantum mechanics to investigate heavy quark mesons is illustrated through 
a study of the charmonium meson.  We consider a limiting form of the QCD potential which is a simple combination of 
the linear and Coulomb potential.  The experimentally determined masses of $J/\psi (1S)$ and $\chi_{c1}(1P)$ are 
reproduced for $m_{c} \approx 1.1 GeV$.  For $\psi(2S)$ we have three different sets of variational parameters and 
to choose the appropriate one we use the leptonic decay width of $\psi(2S)$ and $J/\psi(1S)$.  Finally we use a 
spin-spin interaction to investigate the hyperfine splitting of Charmonium and use it to calculate the mass of 
$\eta_{c}(2S)$.  Our theoretical results agree with the experimentally measured values of $\eta_{c}(2S)$ and 
thereby verifies the usefulness of non-relativistic quantum mechanics in the study of heavy quark meson.  
\end{abstract}

\maketitle

PACS Number(s): 03.65.-w, 14.40.Lb, 12.39.Jh 

\section{Introduction}
\label{Intro}
With the experimental discovery of $J/\psi$ particle [1,2] and the correct identification of it with a bound state of charm quark 
and its anti-particle [3], the study of bound states using non-relativistic quantum mechanics acquired a renewed interest. 
The basic idea is to solve the non-relativistic Schr\"{o}dinger  equation using an interaction potential and fit the bound state 
energy to the mass of $J/\psi$ system.  However, unlike the electromagnetic interaction for which the interaction potential
is known to be the Coulomb potential, the interaction potential for the $c \bar{c}$ system is not known in its completeness. 
A form for the $c \bar{c}$ potential is assumed taking into account that the quarks are permanently confined inside $J/\psi$.
A simple confining potential is a linear potential, $V(r) = kr$, where $k$ is a constant.  

Lucha, Sch\"{o}berl and Gromes [4] have suggested a Coulomb plus linear potential for the $c \bar{c}$ system as 
\begin{equation}
V(r) = -\frac{a}{r} + kr + V_{0}
\label{clpotentialwithconstant}
\end{equation}
The parameters $a$, $k$ and $V_{0}$ have been fitted with $c \bar{c}$ system with the mass of the charmed quark taken 
as $\approx 1.64$ $GeV$ and their values are:
\begin{eqnarray}
a &=& 0.27 (dimensionless),  \nonumber \\
k &=& 0.25 GeV^{2},    \nonumber \\
V_{0} &=& -0.76 Gev.   
\label{potentialparameters}
\end{eqnarray}
Recently Sumino [5] has analysed QCD potential taken as an expectation value of the Wilson loop and showed that in the 
large $N$ limit, the QCD potential behaves like a Coulomb and linear potential.  In view of this we take the $c \bar{c}$
potential to be 
\begin{equation}
V(r) = -\frac{a}{r} + kr
\label{clpotentialnoconstant}
\end{equation}
with $a$ and $k$ as given (\ref{potentialparameters}).  We do not include $V_{0}$ and so consistently treat the mass 
of the charmed quark as a parameter to be determined.  

An exactic analytic solution of the Schr\"{o}dinger  equation with the potential (\ref{clpotentialnoconstant}) is not possible and
so, we use ``Variational method".   Since the potential (\ref{clpotentialnoconstant}) is spherically symmetric, the radial 
wavefunction is guided by known solutions of the $1/r$ potential for bound states.  The motivations of this study are:

\begin{enumerate}
\item To determine the mass of the charmed quark by fitting the $J/\psi$ mass spectrum [6], which is 
\begin{eqnarray}
J/\psi (1S) &=& 3.09687 \pm 0.00004 \; GeV, \nonumber \\
\chi_{c1} (1P) &=& 3.51051 \pm 0.00012 \; GeV, \nonumber \\
\psi(2S) &=& 3.68596 \pm 0.00009 \; GeV. \nonumber
\end{eqnarray}

\item The $J/\psi (1S)$ and $\psi(2S)$ states are spin triplet states.  The recent Belle colloboration [7] observed a new 
charmonium state $\eta_{c} (2S)$, in which the spins of $c\bar{c}$ form a singlet.  The value of the mass for 
$\eta_{c}(1S) = 2.9797 \pm 0.0015$ $GeV$ [6].  These results show a ``hyperfine splitting", as $J/\psi (1S)$ and 
$\eta_{c}(1S)$ have different masses, although both are $c\bar{c}$ bound states.  This cannot be explained by the 
potential model in Eqn. (\ref{clpotentialnoconstant}).  In this work, we introduce a spin-spin interaction and use that 
to explain the hyperfine splitting.  The second motivation is to determine the mass of $\eta_{c}(2S)$ from the results
of hyperfine splitting analysis, consistent with the leptonic decay widths of $J/\psi$ system.  
\end{enumerate}

In Section II, the energies of $J/\psi$ system are calculated for the Coulomb plus Linear potential in variational method. 
For $\psi(2S)$ state, we get three set of values for the variational parameters.  In Section III, the ratio of the leptonic 
decay width of $J/\psi (1S)$ and $\psi(2S)$ are calculated for the three sets.  Comparison with the data, allows us 
to choose one of the three sets.  In Section IV, this set is used to explain the hyperfine splitting after introducing a 
spin-spin contact interaction and the mass of $\eta_{c}(2S)$ is predicted.  This prediction agrees with the recent 
experimental values.  The results are summarized in Section V.

\section{Coulomb plus Linear potential prediction of Charmonium spectrum}
\label{potentialprediction}

The non-relativistic Schr\"{o}dinger equation for the $c\bar{c}$ system is given by 
\begin{equation}
\left( - \frac{\hbar}{2 \mu} \nabla^{2} + V(r) \right) \psi  =  E \psi
\label{nrschrodingerequation}
\end{equation}
where $\mu$ is the reduced mass of the $c\bar{c}$ and is equal to $m_{c}/2$, with $m_{c}$ as the mass of the Charmed 
quark.  Using Spherical polar co-ordinates the Eqns. (\ref{clpotentialnoconstant}) and (\ref{nrschrodingerequation})
are reduced to radial equations.  The radial function $U_{nl}(r)$
\begin{equation}
R_{nl}(r)  =  \frac{U_{nl}(r)}{r} \nonumber
\end{equation}
satisfies 
\begin{equation}
\frac{{\rm d}^{2} U_{nl}}{{\rm d} r^{2}} + \frac{2 \mu}{\hbar^{2}} \left( E_{nl} - Kr + \frac{a}{r} - \frac{\hbar^{2} l (l+1)}{2 \mu r^{2}} \right) U_{nl} = 0
\label{schrodingerequationradial}
\end{equation}
The choice of the trial wave function is dictated by symmetry requirements.  The wave function must vanish at infinity.  The wave function 
must have the angular part of it as $Y_{l}^{m} (\theta, \varphi)$. 

\subsection{$1S$ Bound State}
The trial wave function is taken to be, $\psi_{1s} = R_{1s}(r) Y_{0}^{0}(\theta, \varphi)$ with 
\begin{equation}
R_{1S} = e^{-\lambda r}
\label{radialwf1s}
\end{equation}
where $\lambda$ is the variational parameter.  Then we compute the expectation value of the Hamiltonian as follows:
\begin{equation}
\langle H \rangle = \frac{\langle \psi_{1S} | H | \psi_{1S} \rangle}{\langle \psi_{1S}  \psi_{1S} \rangle}
                            = \frac{\hbar^{2} \lambda^{2}}{2 \mu} - a \lambda + \frac{3 k}{2 \lambda}
\label{variationalequation1S}                            
\end{equation}
On extremising with respect to $\lambda$ we obtain a cubic equation for $\lambda$ as 
\begin{equation}
\frac{\hbar^{2} \lambda^{3}}{\mu} - a \lambda^{2} + \frac{3 k}{2 \lambda} = 0.
\label{minimavariationalequation}
\end{equation}
Using (\ref{minimavariationalequation}) in (\ref{variationalequation1S}), we obtain the minimum value of variational 
energy as 
\begin{equation}
\langle H \rangle_{min} =  \frac{9k}{4 \lambda} -  \frac{a \lambda}{2}
\end{equation}
On account of the choice of (\ref{radialwf1s}), we must use only the real root of the cubit equation 
(\ref{minimavariationalequation}), which is 
\begin{eqnarray}
\lambda = \left[ \frac{3 k \mu}{4 \hbar^{2}} \right]^{\frac{1}{3}}  \left[ \left(1 + \frac{4 \mu^{2} a^{3}}{81 k h^{4}} +
                           \left[1+ \frac{8 \mu^{2} a^{3}}{81 k \hbar^{4}} \right]^{\frac{1}{2}} \right)^{\frac{1}{3}} +
                            \left(1 + \frac{4 \mu^{2} a^{3}}{81 k h^{4}}
                           -  \left[1+ \frac{8 \mu^{2} a^{3}}{81 k \hbar^{4}} \right]^{\frac{1}{2}} \right)^{\frac{1}{3}} \right] 
                           + \frac{\mu a}{3 \hbar^{2}}
\label{realroot1Sstate}
\end{eqnarray}
which expresses $\lambda$ in terms of the potential parameters $a$ and $k$.  We shall then use $\hbar = c = \mu = 1$
system of units. 

Since the mass of charm quark lies between $1.0$ $GeV$ and $1.4$ $GeV$ [6], the values of $\lambda$ and the variational 
energy given by (\ref{realroot1Sstate}) is calculated for $m_{c}$ from $1.0$ $GeV$ and $1.4$ $GeV$.  The variational energy 
is added to the rest mass energy $2m_{c} c^{2}$ after making conversions to CGS units.  The values of the variational parameter
$\lambda$ and the Energy $J/\psi(1S)$ state for values of mass $m_{c}$ are given in Table I.  

\newpage

\begin{table}[!hp]
\caption{Mass of $J/\psi$ $(1S)$ for $m_{c}$ from $1$ $GeV$ to $1.4$ $GeV$} 
\centering  
\begin{tabular}{|c  | c   |c  |} 
\hline                   
Mass of Charm quark    &       Variational Parameter           &  $J/\psi$ $(1S)$ \\
$m_{c}$ $GeV/c^{2}$   &   $\lambda$ (dimensionless)      &           GeV   \\
\hline  
1.0   &  1.2420  & 2.8218 \\
1.1    &  1.1720  & 2.9855 \\
1.2   &  1.1119   & 3.1530  \\
1.3   &  1.0599 & 3.3254  \\
1.4   & 1.0142   & 3.4964  \\
\hline
\end{tabular}
\end{table}

From Table I and using experimental value of $J/\psi(1S) = 3.09687 \pm 0.00004$ $GeV$ [6], 
the Charm quark mass is found to be $m_{c} = 1.165  GeV/c^{2}$. 

\subsection{$1P$ Bound State}
In close analogy with the Coulomb system, the trial wavefunction for the $1P$ state is taken to be, 
$\psi_{1P} = R_{1P}(r) Y_{l}^{m}(\theta, \varphi) $, with 
\begin{equation}
\psi_{1P} =r e^{-\lambda r}
\label{wavefunction1p}
\end{equation}
where the variational parameter $\lambda$ is now for the $1P$-state.  The $1P$ energy  $\langle H \rangle$ is 
found to be,
\begin{equation}
\langle H \rangle = \frac{\langle \psi_{1S} | H | \psi_{1S} \rangle}{\langle \psi_{1S}  \psi_{1S} \rangle}
                            = \frac{\hbar^{2} \lambda^{2}}{2 \mu} - \frac{a \lambda}{2} + \frac{5 k}{2 \lambda}
\label{variationalequation1P}                             
\end{equation}
This is extremized with respect to $\lambda$ and we obtain a cubic equation for $\lambda$, 
\begin{equation}
\frac{\hbar^{2} \lambda^{3}}{\mu} -\frac{ a \lambda^{2} }{2} + \frac{5 k}{2 \lambda} = 0.
\label{minimavariationalequation1P}
\end{equation}
From (\ref{variationalequation1P}) and (\ref{minimavariationalequation1P}) we find, 
\begin{equation}
\langle H \rangle_{min} = \frac{15 k}{4 \lambda} - \frac{a \lambda}{4}
\label{minmumenergy1P}
\end{equation}
The real root of Eqn. (\ref{minmumenergy1P}) of $\lambda$ expressed in terms of $a$ and $k$ is found to be
\begin{eqnarray}
\lambda = \left[ \frac{5 k \mu}{4 \hbar^{2}} \right]^{\frac{1}{3}}  \left[ \left(1 + \frac{ \mu^{2} a^{3}}{270 k h^{4}} +
                           \left[1+ \frac{ \mu^{2} a^{3}}{135 k \hbar^{4}} \right]^{\frac{1}{2}} \right)^{\frac{1}{3}} +
                            \left(1 + \frac{ \mu^{2} a^{3}}{270 k h^{4}}
                           -  \left[1+ \frac{ \mu^{2} a^{3}}{135 k \hbar^{4}} \right]^{\frac{1}{2}} \right)^{\frac{1}{3}} \right] 
                           + \frac{\mu a}{6 \hbar^{2}}.
\label{realroot1Sstate}
\end{eqnarray}

\begin{table}[!hp]
\caption{Mass of $\chi_{c1}$ $(1P)$ for $m_{c}$ from $1$ $GeV$ to $1.4$ $GeV$} 
\centering  
\begin{tabular}{|c  | c   |c  |} 
\hline                   
Mass of Charm quark    &       Variational Parameter           &  $J/\psi$ $(1S)$ \\
$m_{c}$ $GeV/c^{2}$   &   $\lambda$ (dimensionless)      &           GeV   \\
\hline  
1.0   &  1.4022  & 3.2899 \\
1.1    &  1.3186  &  3.4437 \\
1.2   &  1.2469   & 3.6026  \\
1.3   &  1.1844 & 3.7657  \\
1.4   & 1.1295   & 3.9323  \\
\hline
\end{tabular}
\end{table}

From Table II and the experimental value of $\chi_{c1} = 3.51051 \pm 0.00012$  $GeV$ [6], the mass of the 
charmed quark is found to be $m_{c} = 1.1425$ $Gev/c^{2}$ which is in close agreement with the result 
obtained for $1S$ state in II.1. 

\subsection{2S Bound State}
The trial wave function for the $2S$ state is taken to be, $\psi(2S) =  R_{2S} = R_{2S} Y_{0}^{0} (\theta, \varphi)$, with 
\begin{equation}
\psi_{2S} = (b- \lambda r) e^{-\lambda r} 
\label{2Swavefunction}
\end{equation}
where $b$ and $\lambda$ are the variational parameters.  
The $2S$ energy $\langle H \rangle$ is calculated to be 
\begin{equation}
\langle H \rangle = \frac { \frac{\hbar^{2} \lambda^{2}}{8 \mu} (b^{2} -b + 1) 
                            + \frac{3 k}{2 \lambda} (b^{2} -4b +5)  - a \lambda (b^{2} -2b+3/2)}
                            {(b^{2} - 3 b + 3)}
\label{2Senergyexpectationvalue}                            
\end{equation}
Extremization of Eqn. (\ref{2Senergyexpectationvalue}) with respect to $\lambda$ and $b$ leads to 
\begin{equation}
\frac{hbar^{2} \lambda}{\mu} (b^{2} - b + 1) - \frac{3 k}{2 \lambda^{2}} (b^{2} - 4b +5) - a (b^{2} - 2b + 3/2) =0
\label{minimaequationI}
\end{equation}
\begin{equation}
\frac{hbar^{2} \lambda}{\mu} (-b^{2} + 2b) + \frac{3 k}{2 \lambda^{2}} - a ( -b^{2} + 3b - 3/2) = 0
\label{minimaequationII}
\end{equation}
On simplifying (\ref{2Senergyexpectationvalue}) using (\ref{minimaequationI}) and (\ref{minimaequationII}) the minimum 
value of the variational energy $\langle H \rangle_{min}$ is found to be 
\begin{equation}
\langle H \rangle_{min} = \frac{ \frac{9k}{4 \lambda} (b^{2} - 4b + 5) - \frac{a \lambda}{2} (b^{2} -2b + 3/2) }{(b^{2} -3b + 3)}
\label{Hminima2S}
\end{equation}
From (\ref{minimaequationI}) and (\ref{minimaequationII}), we have 
\begin{equation}
b = \frac{3k - \lambda^{3}}{\lambda^{3} - a \lambda^{2}}
\label{parameterb}
\end{equation}
We have to first determine $\lambda$ and $b$ consistently satisfying (\ref{minimaequationI}) and (\ref{minimaequationII}). 
We used two independent methods for this analysis.  First, $\lambda$ is assumed to be in the range $0.1$ and $2.0$ through a 
trial and error method.  For each value of $\lambda$, Eqn. (\ref{parameterb}) determines b.  Then (\ref{minimaequationI} 
and (\ref{minimaequationII}) are required to be satisfied.  This happens for three sets of values of $(\lambda, b)$.  In 
the second method, a 'random number generator' is used to choose $\lambda$.  Determining $b$ from Eqn. (\ref{parameterb}), 
the random number generator finds $\lambda$ such that equations (\ref{minimaequationI}) and (\ref{minimaequationII}) are 
required to be satisfied.  This gave three sets of values for $(\lambda, b)$ which agrees with the first method.  
Among the three sets of values, two of them have both $\lambda$ and $b$ positive, meanwhile in the third set while $\lambda$
is still positive, $b$ was $-ve$.  Using all the three sets of results, the $\psi(2S)$ energy is calculated.  

Set 1:  ($\lambda > 0$ ; $b>0$)

\begin{table}[!hp]
\caption{Mass of $\psi(2S)$ for $m_{c}$ from $1$ $GeV$ to $1.4$ $GeV$} 
\centering  
\begin{tabular}{|c  |c   |c  |c |} 
\hline                   
Mass of Charm quark    &       Variational Parameter         &   Variational Parameter       &  $J/\psi$ $(1S)$ \\
$m_{c}$ $GeV/c^{2}$   &   $\lambda$ (dimensionless)    &  $b$ (dimensionless)          &           GeV   \\
\hline  
1.0   &  1.1322  & 1.4012   & 3.7100 \\
1.1    &  1.0663 & 1.3992  &  3.8485 \\
1.2   &  1.0095 & 1.3974  & 3.9927  \\
1.3   &  0.9605 & 1.3956  & 4.1439 \\
1.4   &  0.9171  & 1.3938  & 4.2986  \\
\hline
\end{tabular}
\end{table}
Using the experimental value of $\psi(2S) = 3.68596 \pm 0.00009 GeV$ [6] and Table III the values of $b$ and $\lambda$ 
are 
\begin{eqnarray}
b &=& 1.4012  \rm{dimensionless}  \nonumber \\
\lambda &=& 1.1322 \rm{dimensionless}  \nonumber \\
m_{c} &=& 1.0 GeV/c^{2} \nonumber
\end{eqnarray}

\newpage

Set 2:  ($\lambda > 0$ ; $b>0$)

\begin{table}[!hp]
\caption{Mass of $\psi(2S)$ for $m_{c}$ from $1$ $GeV$ to $1.6$ $GeV$} 
\centering  
\begin{tabular}{|c  |c   |c  |c |} 
\hline                   
Mass of Charm quark    &       Variational Parameter         &   Variational Parameter       &  $J/\psi$ $(1S)$ \\
$m_{c}$ $GeV/c^{2}$   &   $\lambda$ (dimensionless)    &  $b$ (dimensionless)          &           GeV   \\
\hline  
1.0   &  0.9147  & 4.1423   & 2.8063 \\
1.1    &  0.8636 & 4.1466   & 2.9704 \\
1.2   &  0.8196  & 4.1509   & 3.1384 \\
1.3   &  0.7815  & 4.1551    & 3.3094 \\
1.4   &  0.7480 & 4.1594    & 3.4831  \\
1.5   &  0.7183  & 4.1634    & 3.6587  \\
1.6   &  0.6917  & 4.1678    & 3.8363  \\
\hline
\end{tabular}
\end{table}
From the above table and using the experimental value of $\psi(2S) = 3.68596 \pm 0.00009 GeV$ [6] the mass of the charm 
quark was found to be $1.515 GeV/c^{2}$ and the corresponding values of the variational parameters are $\lambda = 0.714078$
dimensionless and $b = 4.164253$ dimensionless.  

Set 3:  ($\lambda > 0$ ; $b<0$)

\begin{table}[!ht]
\caption{Mass of $\psi(2S)$ for $m_{c}$ from $1$ $GeV$ to $1.6$ $GeV$} 
\centering  
\begin{tabular}{|c  |c   |c  |c |} 
\hline                   
Mass of Charm quark    &       Variational Parameter         &   Variational Parameter       &  $J/\psi$ $(1S)$ \\
$m_{c}$ $GeV/c^{2}$   &   $\lambda$ (dimensionless)    &  $b$ (dimensionless)          &           GeV   \\
\hline  
1.0   &  1.9392   &  -0.6839   & 2.7831 \\
1.1    &  1.8267  & -0.6962     & 2.9049 \\
1.2   &  1.7298  & -0.7083     &  3.1172  \\
1.3   &  1.6461  & -0.7202     & 3.2895 \\
1.4   &  1.5720  & -0.7319     & 3.4638  \\
1.5   &  1.5062  & -0.7435    &  3.6400 \\
1.6   &  1.4481  & -0.7548    &  3.8189 \\
\hline
\end{tabular}
\end{table}
Using the experimental value of $\psi(2S) = 3.68596 \pm 0.00009 GeV$ [6] and Table V, we find the mass of charm 
quark as $1.5275  GeV/c^{2}$ and the corresponding values of the variational parameters are, 
\begin{eqnarray}
\lambda &=& 1.48996  \rm{dimensionless},  \\
b &=& -0.74657 \rm{dimensionless}.  
\end{eqnarray}

Thus, we have three set of results for $\psi(2S)$ state.  In order to choose one among the three sets we calculate the 
``Leptonic Decay Width" of $J/\psi(1S)$ and $\psi(2S)$ states and compare with their experimental values.  

\section{Leptonic decay width}
\label{leptonic}
The leptonic decay width of a typical $(nS)$ state is given by [8]
\begin{equation}
\Gamma(nS \rightarrow e^{+} e^{-})  = \frac{4 \alpha^{2} e_{q}^{2}}{M_{n}^{2}}   |R_{n}(0)|^{2} \gamma_{n}
\label{leptonicdecaywidth}
\end{equation}
where $\alpha$ is the QCD coupling strength, $e_{q}$ is the charge of the quark, $M_{n}$ of the $n^{th}$ state meson, 
$R_{n}(0)$ is the value of the $n^{th}$ state radial wavefunction at the origin and $\gamma_{n}$ is the QCD correction
factor.  

For the $2S$ and $1S$ states of $c\bar{c}$ meson, we have 
\begin{equation}
R = \frac{\Gamma^{\psi} (2S \rightarrow e^{+} e^{-})}{\Gamma^{\psi} (1S \rightarrow e^{+} e^{-})}
   = \frac{|R_{2S} (0) |^{2}} {|R_{1S}(0)|^{2}} \times \frac{M^{2}_{1s}}{M^{2}_{2s}}
       \times \frac{\gamma_{2}}{\gamma_{1}}
\label{leptonicdecaywidthccbar}
\end{equation}
where we have assumed the fact that $\alpha(1S)$ and $\alpha(2S)$ are nearly equal to one another and 
$\gamma_{1} =0.581$ and $\gamma_{2} = 0.606$ [8].   From particle physics data group [6], the experimental 
values of the leptonic decay width for $1S$ and $2S$ states give the ratio
\begin{equation}
R = \frac{\Gamma^{\psi} (2S \rightarrow e^{+} e^{-})}{\Gamma^{\psi} (1S \rightarrow e^{+} e^{-})} = 0.4163498.
\end{equation}
For the $2S$ state we have three set of results and so consequently three set of values for $\lambda$ and $b$. 

We use Eqn. (\ref{leptonicdecaywidthccbar}) to calculate the ratio between the $2S$ leptonic decay width and 
$1S$ leptonic decay width of $c \bar{c}$, using the wave function for $J/\psi(1S)$ and $\psi(2S)$ after 
normalization with $\lambda$ and $b$ from the three sets as shown in the table below: 
\begin{table}[!ht]
\caption{Values of R for the three sets in Section II.C} 
\vspace{0.1cm}
\centering  
\begin{tabular}{|c  |c   |} 
\hline              
Set     &    R dimensionless \\
\hline 
1         &     1.9146  \\
2        &      0.4107  \\
3        &      0.1623  \\
\hline 
\end{tabular}
\end{table}
From the Leptonic decay width ratios for the three sets, we find that the result for the second set agrees with the 
experimental value.  So we choose the mass value corresponding to second set of result.  The mass of charm quark 
is $m_{c} = 1.515 GeV/c^{2}$.  

\section{Hyperfine splitting}
In order to provide further check on the second set for the $\psi(2S)$ state, we consider the hyperfine splitting of charmonium 
levels.  In the $c\bar{c}$ system there are two types of states; the $J/\psi (1S)$ and the $\psi(2S)$ are spin triplet states
and the $\eta_{c}$ states are the spin singlet states.  Their masses are not the same.  This cannot be explained by the 
potential given in Eqn. (\ref{clpotentialnoconstant}).  We therefore introduce a spin dependent interaction of the form 
$\alpha  \vec{S_{1}} \cdot \vec{S_{2}} \delta^{3}(\vec{r})$ in the Hamiltonian where $\alpha$ is taken to be a constant 
for simplicity [8].  So the Hamiltonian is of the form, 
\begin{equation}
H = \frac{p^{2}}{2 \mu} + k r - \frac{a}{r} + \alpha  \vec{S_{1}} \cdot \vec{S_{2}} \delta^{3}(\vec{r})
\label{Hamiltonianwithspinspininteraction}
\end{equation}
The expectation of the Hamiltonian in Eqn. (\ref{Hamiltonianwithspinspininteraction}) is 
\begin{equation}
E = E_{0} + \alpha  \langle \vec{S_{1}} \cdot \vec{S_{2}} \rangle \langle \delta^{3}(\vec{r}) \rangle . 
\label{Energyspinspininteraction}
\end{equation}
where $E_{0}$ represents the spin independent part of the energy.   For the spin triplet state, 
\begin{equation}
 \langle \vec{S_{1}} \cdot \vec{S_{2}} \rangle = \frac{1}{4}  \nonumber
\end{equation}
For the spin singlet state
\begin{equation}
 \langle \vec{S_{1}} \cdot \vec{S_{2}} \rangle = - \frac{3}{4}  \nonumber.
\end{equation}
Then 
\begin{eqnarray}
E_{1S} (J/\psi)      &=&  E_{0} + \frac{\alpha}{4} [R_{1S}(J/\psi)_{r=0}]^{2}  \\
E_{1S} (\eta_{c})  &=&  E_{0} - \frac{3 \alpha}{4} [R_{1S}(\eta_{c})_{r=0}]^{2}. 
\end{eqnarray}
Similarly for $2S$ state, 
\begin{eqnarray}
E_{2S}(\psi)  &=& E_{0} + \frac{\alpha}{4} [R_{2S}(J/\psi)_{r=0}]^{2}   \\ 
E_{2S}(\psi)  &=&  E_{0} - \frac{3 \alpha}{4} [R_{2S}(\eta_{c})_{r=0}]^{2}. 
\end{eqnarray}
From the above set of relations we get
\begin{eqnarray}
E_{1S}(J/\psi) - E_{1S}(\eta_{c}) &=& 2 \alpha [R_{1S} (r = 0)]^{2} = \Delta H (1S)  \nonumber \\
E_{2S}(J/\psi) - E_{2S}(\eta_{c}) &=& 2 \alpha [R_{2S} (r = 0)]^{2} = \Delta H (2S)
\label{energydiff}
\end{eqnarray}
\begin{equation}
\frac{\Delta H (1S)}{\Delta H (2S)} \approx \frac{[R_{2S}(r=0)]^{2}}{[R_{1S}(r=0)]^{2}}.
\label{energydiffratio}
\end{equation}
In \label{energydiffratio} we assumed that $\alpha$ does not change much for $1S$ and $2S$ states.  
The ratio of $\frac{\Delta H (1S)}{\Delta H (2S)}$ for our set of values i.e., the second set is 
\begin{equation}
frac{\Delta H (1S)}{\Delta H (2S)}  = 0.5578 {\rm dimensionless}
\label{ratioexperimental}
\end{equation}
From the particle data group [6] we know that $\Delta (1S) = 117.2 MeV$.  Then Eqn. (\ref{ratioexperimental})
gives 
\begin{equation}
\Delta H (2S) = 65.37 MeV
\end{equation}
Since, $\Delta H(2S) = E_{2S}(\psi) - E_{2S}(\eta_{c})$ our calculation gives
\begin{equation}
E_{2S}(\eta_{c})  = 3.62059 GeV. 
\end{equation} 
The recent observation of Charmonium $\eta_{c}(2S)$ by the Belle Collaboration yielded the following values 
for its mass
\begin{equation}
M(\eta_{c} (2S))  =  3.654 (14) GeV  [9]
M(\eta_{c} (2S))  = 3.622  (12) Gev  [7]. 
\end{equation}
The value of $E_{2S}(\eta_{c})$ obtained from our calculation agrees with the value of $M(\eta_{c}(2S))$ 
experimentally obtained.  

\section{Conclusion}
\label{conclusion}
We have used a Coulomb plus linear potential to explain the mass of $c \bar{c}$ spectrum, employing the 
variational method.  $J/\psi(1S)$ and $\chi_{c1}(1P)$ masses are correctly reproduced for $m_{c} \approx 1.1 GeV$. 
$\psi(2S)$ state is explained with three set of values for the variational parameters.  In order to find the correct
set, we calculate leptonic decay width of $\psi(2S)$  and $J/\psi(1S)$.  By comparing with experiment, the 
second set is found to explain both the mass and leptonic decay width.  This set is used in conjunction with 
spin-spin contact interaction to explain the hyperfine splitting of Charmonium.  As a consequence, we are able
to calculate the mass of $\eta_{c} (2S)$ as $3.6026 GeV$ which agrees reasonably well with the experiment. 
This study indicates the ability of non-relativisitic approach in explaining the $c \bar{c}$ system.  

%
%
%

\section{References}

[1]  J.J. Aubert et. al., {\it Phys. Rev. Lett} {\bf 33}, 1404 (1974).   \\

[2] J.E. Augustine et. al.,  {\it Phys. Rev. Lett} {\bf 33}, 1406 (1974).   \\

[3] G. Goldhaber et. al., {\it Phys. Rev. Lett} {\bf 37}, 255 (1976).   \\

[4] W. Lucha, F.F. Sch\"{o}berl and D. Gromes, {\it Physics Reports} {\bf 200} 126 (1991).   \\

[5] Y. Sumino {\it arXiv No: hep-ph/0303120}. \\

[6] Particle Data Group K. Hagiwara et. al., {\it Phys. Rev. D} {\bf 66}, 010001 (2002).  \\

[7] Belle Collaboration, K. Abe et. al., {\it Phys. Rev. Lett} {\bf 89}, 142001 (2002). \\

[8] A.M. Badalian and B.L.G. Bakker  	 {\it arXiv No: hep-ph/0302200}.  \\

[9] Belle Collaboration, S.K. Choi et. al., {\it Phys. Rev. Lett} {\bf 89} 102001 (2002). 
\end{document}